\documentclass{article}
\usepackage{ijcai25}

\usepackage{times}
\usepackage{soul}
\usepackage{url}
\usepackage[hidelinks]{hyperref}
\usepackage[utf8]{inputenc}
\usepackage[small]{caption}
\usepackage{graphicx}
\usepackage{amsmath}
\usepackage{amsthm}
\usepackage{natbib}
\usepackage{booktabs}
\usepackage{algorithm}
\usepackage{algorithmic}
\usepackage[switch]{lineno}
\usepackage{bm}
\usepackage{amsfonts}
\usepackage{amssymb}
\usepackage{array}
\usepackage{xcolor}
\usepackage{multirow}
\usepackage{colortbl} 

\title{Why Regression? Classification with Stock Component Fusion \\ for Stock Index Predictions}
\title{Why Regression? Predicting Index Price with Stock Component Fusion and Binary Encoding Classification}
\title{Why Regression? Binary Encoding Classification Brings Confidence to \\ Stock Market Index Price Prediction}

\author{
Junzhe Jiang$^{1,}$\thanks{Equal contribution, alphabetical order}~, Chang Yang$^{1,}$\footnotemark[1]~,  Xinrun Wang$^{2,}$\footnotemark[2],\textbf{Bo Li$^{1}$}
\\
\emails
$^1$The Hong Kong Polytechnic University, $^2$Singapore Management University
\texttt{junzhe.jiang@connect.polyu.hk, xrwang@smu.edu.sg}}

\begin{document}

\maketitle

\begin{abstract}

Stock market indices serve as fundamental market measurement that quantify systematic market dynamics.  However, accurate index price prediction remains challenging, primarily because existing approaches treat indices as isolated time series and frame the prediction as a simple regression task. These methods fail to capture indices' inherent nature as aggregations of constituent stocks with complex, time-varying interdependencies. To address these limitations, we propose \textsc{Cubic}, a novel end-to-end framework that explicitly models the adaptive fusion of constituent stocks for index price prediction.  
Our main contributions are threefold.
i) Fusion in the latent space: we introduce the fusion mechanism over the latent embedding of the stocks to extract the information from the vast number of stocks. 
ii) Binary encoding classification: since regression tasks are challenging due to continuous value estimation, we reformulate the regression into the classification task, where the target value is converted to binary and we optimize the prediction of the value of each digit with cross-entropy loss. 
iii) Confidence-guided prediction and trading: we introduce the regularization loss to address market prediction uncertainty for the index prediction and design the rule-based trading policies based on the confidence.  
Extensive experiments across multiple stock markets and indices demonstrate that \textsc{Cubic} consistently outperforms state-of-the-art baselines in stock index prediction tasks, achieving superior performance on both forecasting accuracy metrics and downstream trading profitability.    


\end{abstract}

\section{Introduction}
The prediction of stock market indices has long been an intriguing topic in financial research, driven by the potential for stable profits through index option trading \cite{chalvatzis2020high}. Stock market indices, such as the Dow Jones Industrial Average (DJIA), measure market value through selected representative stocks \cite{lin2021forecasting}. Accurate index predictions are crucial for modern financial markets, underpinning portfolio management, derivatives trading, and various financial products that collectively represent trillions in trading volume \cite{coles2022index}.

In recent years, with the rapid development of artificial intelligence and increasing market complexity, there has been a surge of interest in applying deep learning models to stock price prediction. Various architectures such as transformers \cite{vaswani2017attention}, multilayer perceptrons (MLP) \cite{kara2011predicting}, and long-short-term memory networks (LSTMs) \cite{graves2012long} have been extensively explored in financial forecasting tasks. These models have demonstrated remarkable capabilities in capturing complex patterns and temporal dependencies inherent in financial time series data, with the aim of improving the accuracy of stock index forecasting. Recent advances in deep learning architectures have led to significant improvements in stock prediction performance, such as an attribute-driven graph attention network \cite{cheng2021modeling} that models the momentum spillover effect and effectively captures complex relationships among stocks. The market-guided stock transformer (MASTER) \cite{li2024master} enhanced the traditional Transformer architecture by incorporating market information to guide the attention mechanism, while StockMixer \cite{fan2024stockmixer} introduced an innovative MLP-based architecture that successfully captures nonlinear relationships in financial data. Deep Transformer architectures \cite{wang2022stock} have demonstrated superior capabilities in handling long-sequence time series data, and attention-enhanced GRU models \cite{sethia2019application} have proven effective in capturing temporal dependencies. Additionally, hybrid approaches combining traditional signal processing with deep learning, such as the CEEMDAN-LSTM model \cite{lin2021forecasting}, have shown improved prediction performance, while comprehensive studies on LSTM applications \cite{fischer2018deep} have validated their effectiveness in handling temporal dependencies and non-linear patterns in financial data.

Despite these advances, existing methods face two major limitations~\cite{torres2021deep}. First, predicting stock indices is inherently more complex than predicting individual stocks~\cite{wang2021stock}, as indices are composed of multiple stocks with different weights. This composition not only increases the number of variables and factors to consider but also leads to the \textit{curse of dimensionality} problem. Moreover, indices typically include stocks from various industries with complex interactions and correlations, where changes in one industry can trigger chain reactions in others, further complicating the prediction task. Second, the prevalent approach of formulating price prediction as a regression problem faces significant challenges in stock index prediction. The high randomness and volatility of financial markets~\cite{liang2022forecasting} make it difficult for regression models to accurately fit price change trends. These models assume continuous and smooth price movements, failing to effectively capture sharp fluctuations and sudden market events. Their sensitivity to outliers often results in suboptimal prediction performance, suggesting that regression-based approaches may not be the most suitable choice for this task.


To address the above challenges, 
we propose a novel framework, i.e., \textbf{C}omponent f\textbf{U}sion and \textbf{B}inary encoding class\textbf{I}fication with \textbf{C}onfidence (\textsc{Cubic}), for the prediction of stock index. Our framework serves as a general-purpose enhancement that can work with various deep learning architectures to improve prediction and trading performance.
The four main contributions of \textsc{Cubic} are:
\begin{itemize}
    \item We present a novel formulation that transforms the prediction of stock index from time series regression to binary encoding classification through target value binarization, enabling robust prediction with quantification of inherent uncertainty.

    \item We employ fusion in the latent space to tackle the issue of fixed stock weights when calculating the stock index and capturing nonlinear relationships between stocks.
    
    
    \item We introduce confidence-guided prediction and trading that leverages classification-based probability outputs as natural confidence measures and incorporates specialized regularization loss to enhance prediction reliability and guide trading decisions.

    \item To the best of our knowledge, \textsc{Cubic} is the first to consider stock component fusion and leverage binary encoding classification for index price prediction. Extensive experiments on three representative market indices demonstrate its consistent superiority in both prediction and trading performance, which may motivate broader applications in financial markets.

\end{itemize}



\textbf{Related Work.} Stock market index prediction evolved from statistical approaches such as ARMA \cite{zhang2024hybrid,pokou2024hybridization} to advanced deep learning architectures. The transformer \cite{vaswani2017attention} and variants have become dominant in this field, and works such as \cite{wang2022adaptive,zhang2018stock} demonstrate their effectiveness in capturing market dynamics. The encoder-decoder architecture and multi-head attention mechanism have proven particularly adept at handling non-linear characteristics of financial markets, shown in applications, including relation-aware portfolio learning \cite{xu2021relation} and hierarchical market prediction \cite{ding2020hierarchical}. LSTM-based models \cite{hochreiter1997long,ye2023self} continue to show strong performance in capturing temporal dependencies, while hybrid approaches combining traditional and deep learning methods \cite{kamara2022ensemble} have emerged to improve robustness. Recent works advanced the field through innovations such as period correlation mechanisms \cite{tao2024series}, heterogeneous information fusion \cite{wang2022graph}, and attention mechanisms \cite{liu2024enhanced}. In particular, there has been a growing trend in reformulating regression problems as classification tasks, with \cite{stewart2023regression} providing theoretical evidence for advantages of this approach over traditional MSE-based regression, particularly in neural networks. This classification-based paradigm has shown promise in various financial applications, from capturing market volatility patterns \cite{kyoung2019performance} to improving deep reinforcement learning for portfolio optimization \cite{farebrother2024stop,wang2023categorical}. The effectiveness of discrete classification frameworks has been validated by works such as \cite{vzlivcar2018discrete}, which introduces novel discretization schemes for financial data to enhance model robustness.

\section{Problem Statement}


Let $[N]$ denote the set of constituent stocks in the target market index. For each stock $i \in [N]$, we extract $M$ technical indicators at each time step $t \in [1,\tau]$, yielding feature vectors $x_{i,t} \in \mathbb{R}^M$. Based on established research \cite{huynh2023efficient}, we incorporate 16 technical indicators categorized into three groups: Trend indicators for directional movements, Oscillator indicators for momentum and reversal signals, and Volatility indicators for risk levels, which collectively capture essential market dynamics. The detailed mathematical formulations are presented in Table~\ref{tab:indicator} in Appendix~\ref{appendix_indicator}, with a concise overview in Table~\ref{tab:indicator_summary}. In line with \cite{fan2024stockmixer}, we define our prediction target as the market return: $y_t = \frac{I_{t+1} - I_t}{I_t}$, where $I_t$ represents the market index value at time $t$. To ensure numerical stability, we apply standardization to obtain $\hat{y}_t = \text{standardize}(y_t)$. The market index prediction task can be formulated as: given the technical indicators $\{x_{i,s}\}_{i\in[N], s\in[t-\tau+1,t]}$ of constituent stocks over $\tau$ time steps, predict the normalized next-day market return $\hat{y}_t$.

\begin{table}[ht]
    \centering
    \caption{Summary of the technical indicators used.}
    \begin{tabular}{c|c}
    \toprule[1.5pt]
    \textbf{Type} & \textbf{Indicators}\\
    \midrule[1pt]
    \multirow{3}{*}{Trend} & Arithmetic Ratio, Open, Close, \\
    & Close SMA, Volume SMA, \\ 
    &Close EMA, Volume EMA, ADX  \\
    \midrule
    Oscillator & RSI, MACD, MACD Signal, K, MFI\\
    \midrule
    Volatility & ATR, BB Middle, OBV\\

    \bottomrule[1.5pt]
    \end{tabular}
    
    \label{tab:indicator_summary}
\end{table}

\section{\textsc{Cubic}}

\begin{figure*}
    \centering
    \includegraphics[width=1\textwidth]{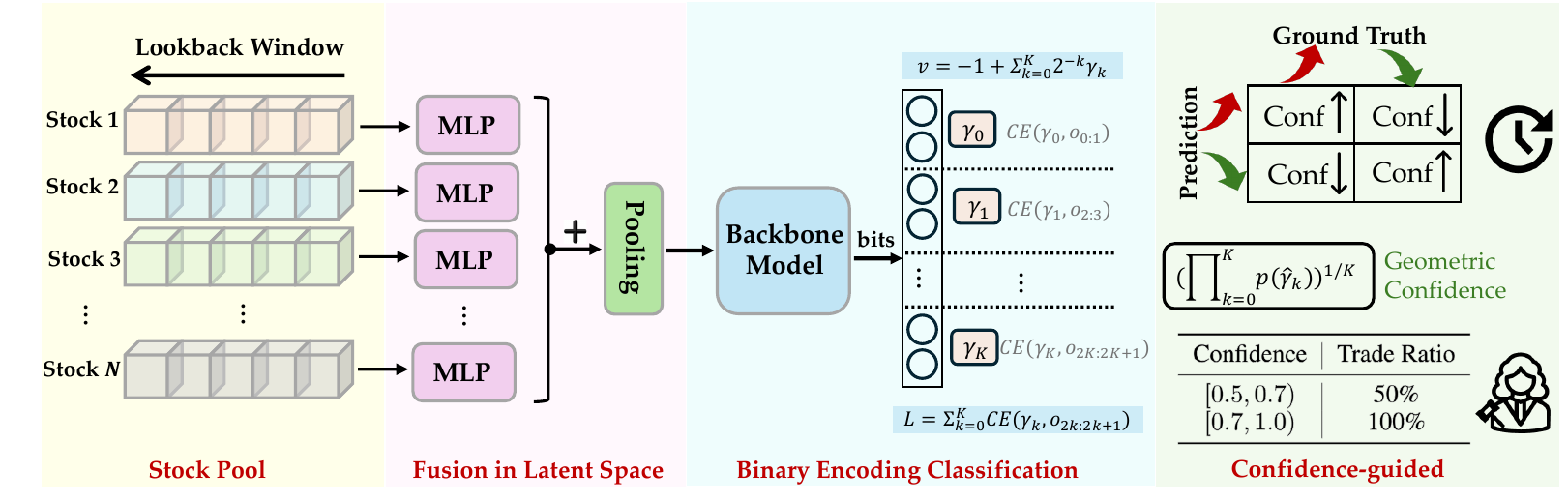}
    \caption{LLM applications, data, code and benchmarks, and challenges and opportunities in quant finance}
    \label{fig:overview}
\end{figure*}

This section introduces the \textsc{Cubic} framework, as shown in Figure~\ref{fig:overview}. Specifically, \textsc{Cubic} first embed the stock indicators into the latent space and perform fusion using pooling techniques. 
Then, we introduce binary encoding classification for stock index price prediction, where the continuous target value is discretized and transformed into a binary representation.
Finally, \textsc{Cubic} leverages confidence to construct the regularization loss for accurate prediction and trading.

\subsection{Fusion in Latent Space}

A market index typically comprises hundreds of constituent stocks, for example, CSI 100 consists of 100 stocks, posing significant challenges for deep learning models to effectively extract meaningful patterns from such high-dimensional data. This high dimensionality can lead to overfitting issues where models capture noise rather than the market dynamics that drive stock behaviors. Moreover, the large-scale nature of constituent stocks substantially increases computational complexity, impeding efficient model training and deployment. Therefore, we propose fusion in latent space to effectively distill and integrate information across constituent stocks.

\paragraph{Feature Embedding of Stocks.} 

Given the technical indicators (e.g., arithmetic ratio), we introduce an embedding model to project stock indicators into a latent space of 32 dimensions. This embedding mechanism learns to capture complex cross-stock patterns while reducing dimensionality for efficient computation. Specifically, we employ a Multi-Layer Perceptron (MLP) for each stock to obtain its embedding:
\begin{equation}
    \bm{e}^{i} = \text{EMB}(\bm{x}^{i}), i\in[N],
\end{equation}
where $\bm{x}^{i}$ denotes the input features of stock $i$, and $\bm{e}^{i}$ represents its learned embedding. The embedding model $\text{EMB}(\cdot)$ is implemented as an MLP to capture non-linear relationships in high-dimensional stock features.

\paragraph{Pooling.} 
Given the embeddings of constituent stocks, a straightforward approach would be to concatenate all stock embeddings into a flat vector. However, this becomes computationally intractable as the number of stocks increases - for instance, with 1000 stocks and 32-dimensional embeddings per stock, this would result in a 32000-dimensional vector. To achieve efficient information aggregation while preserving crucial market signals, we introduce a multi-head pooling mechanism that combines three complementary operations:
\begin{itemize}
\item The max pooling extracts the most salient features across stocks by selecting the maximum value: $\hat{\bm{e}}=\langle\hat{e}{j}\rangle$, where $\hat{e}{j}=\max_{i\in[N]}{e_{j}^{i}}$. This operation helps capture extreme market movements and dominant patterns.

\item The mean pooling computes the average value between stocks: $\hat{\bm{e}}=\langle\hat{e}_{j}\rangle$, where $\hat{e}_{j}=\sum_{i\in[N]}\{e_{j}^{i}\}/N$. This aggregation preserves the overall trend of the market and the collective behavior of stocks.

\item The min pooling captures the lower bounds of feature distributions: $\hat{\bm{e}}=\langle\hat{e}_{j}\rangle$, where $\hat{e}_{j}=\min_{i\in[N]}\{e_{j}^{i}\}$. This helps identify market downturns and potential risks.

\end{itemize}
The combination of these three pooling operations ($\hat{\bm{e}}=\langle\hat{e}{\text{max}}, \hat{e}{\text{mean}}, \hat{e}_{\text{min}}\rangle$) provides a comprehensive view of the market by capturing diverse aspects of stock behavior: peak signals through maximum pooling, average trends through mean pooling, and bottom signals through min pooling. This multihead design addresses the challenge of dimension while allowing the model to learn from multiple market perspectives in a computationally efficient manner.

\subsection{Binary Encoding Classification}

Accurate price prediction through regression poses significant challenges in financial markets. Traditional regression with mean-squared error (MSE) loss often struggles with vanishing gradients when the loss becomes small, making it difficult for the model to capture subtle but crucial price movements. Additionally, regression models tend to be particularly sensitive to noise and outliers in volatile financial data, potentially leading to unstable and unreliable predictions.

Therefore, we reformulate the price prediction task by encoding continuous price values into binary representations. Specifically, \textsc{Cubic} first converts each price value into its binary format, treats each binary digit as a classification target, and then reconstructs the original price from the predicted binary digits. This binary encoding strategy transforms a challenging regression problem into multiple simpler binary classification tasks, providing more stable gradients through cross-entropy loss while maintaining the model's ability to predict precise price values.

\paragraph{Binary Encoding.} For the target value $v\in[-1, 1]$, we represent it with binary:
\begin{align}
    v = -1 + \sum\nolimits_{k=0}^{K} \gamma_{k}\cdot 2^{-k}, \gamma_{k}\in\{0, 1\}
\end{align}
where $K=15$ is chosen to achieve a precision of 0.0001, and $\gamma_{k}$ represents each bit in the binary representation.

\paragraph{Prediction via Classification.} The model outputs $2K$ dimensions, with each pair of dimensions predicting one binary digit. We propose a weighted cross-entropy loss to emphasize the hierarchical nature of binary encoding. Given the binary encoding $\bm{\gamma}$ and model output $\bm{o}$, the loss is:
\begin{equation}
    L_{CE}(\bm{\gamma},\bm{o})=\sum\nolimits_{k=0}^{K}w_k\texttt{CE}(\gamma_{k}, \bm{o}_{2k:2k+1})
\end{equation}
where $\texttt{CE}(\cdot)$ is the cross-entropy loss and $w_k$ denotes the position-dependent weight. The binary decomposition induces a multi-resolution representation of target values, with each bit position corresponding to different magnitudes in the numerical space. Such multi-scale learning, combined with the stable gradients from cross-entropy loss and position-dependent weighting, enables more effective learning from limited financial data compared to direct regression.

\subsection{Confidence-guided Prediction and Trading}

Introducing classification into regression tasks has a natural advantage that predicted outputs are inherently confident. Specifically, when the output values of the two classes exhibit a large disparity, the confidence of the model in the class with higher value will increase. Therefore, we leverage this confidence and design the regularization loss to guide the predictions and inform trading decisions. Specifically, for each binary digit $k$, given the model output probabilities for both classes, we determine the predicted bit value as:
\begin{equation}
    \hat{\gamma}_{k} = \arg\max\nolimits_{c\in\{0, 1\}} \{p(c)\}
\end{equation}
where $p(c)$ represents the model's predicted probability for class $c$. We propose two variants of geometric confidence (GC) to capture different aspects of prediction reliability:

\paragraph{Mean Confidence.} The geometric confidence (GC) measure is defined as: $GC_{mean}\left(\prod_{k=0}^{K}p(\hat{\gamma}_{k})\right)^{1/K}$, where $p(\hat{\gamma}_{k})$ represents the model's predicted probability for the chosen bit value $\hat{\gamma}_{k}$ at position $k$. This geometric mean of probabilities across all binary digits evaluates the model's comprehensive understanding of price patterns across all scales.

\paragraph{Trend Confidence.} To capture confidence in directional movement prediction, we define: $GC_{trend} = p(\hat{\gamma}_{0})$, which examines confidence of the most significant bit. This measure reflects the model's certainty in predicting price trend direction, crucial for directional trading decisions.

These complementary confidence measures enable more nuanced trading strategies by considering both the overall prediction reliability and the directional movement confidence. A high $GC_{mean}$ indicates consistent confidence across all price scales, while a high $GC_{trend}$ suggests strong conviction in the predicted price trend direction.

\paragraph{Confidence-guided Prediction.} 

To enhance the model's predictive reliability and capability, we leverage geometric confidence to guide prediction behavior through a confidence-aware regularization mechanism. Specifically, when the model correctly identifies the directional indicator corresponding to the sign of the predicted trajectory, our goal is to maximize confidence in the predicted value. Conversely, when the model's predicted trend is incorrect, we tend to minimize the confidence in the predicted value. As $p(\gamma_{0})$ can determine the trend of values predicted by the model, we formulate the confidence-guided regularization loss:

\begin{equation}
    L_{\text{Conf}} = (1-2\cdot \mathbb{I}[p(\gamma_{0}) > p(1-\gamma_{0})]) \cdot GC.
\end{equation}
The total prediction loss is defined as: $L_{\text{pred}} = L_{CE} +  L_{\text{Conf}}$ This adaptive regularization term ensures that the model learns to express high confidence only when predictions are reliable, improving the robustness of the prediction system.

\paragraph{Confidence-guided Trading.} 
\label{app:confidence-trading}
The defined geometric confidence serves as a crucial indicator for optimizing trading decisions. Specifically, we implement a dynamic position sizing strategy based on the confidence levels: when the model signals an upward price movement with moderate confidence between 0.5 and 0.7, we adopt a conservative approach by allocating 50\% of the available position size. For high-confidence predictions ranging from 0.7 to 1, we execute full position trades to maximize potential returns. This adaptive allocation mechanism applies symmetrically to downward price predictions. The confidence-guided trading framework enables sophisticated risk management and trade execution optimization, facilitating the construction of robust and adaptive trading strategies that potentially enhance risk-adjusted returns.

\section{Experiments}
To evaluate the effectiveness of CUBIC framework and plug-and-play components, we conduct experiments on three major indices, selected as representative benchmarks of their markets. This design validates that CUBIC can consistently improve performance in various market conditions and model architectures. In this section, we first introduce the experimental setup and then answer three research questions (RQs). \textbf{RQ1}: Can CUBIC'S fusion in latent space and binary encoding improve model performance compared to direct concatenation and standard regression? \textbf{RQ2}: Can CUBIC'S confidence-guided prediction and strategy achieve superior prediction accuracy and trading performance through selective trading decisions? \textbf{RQ3}: How well does CUBIC demonstrate consistent performance improvements across different markets, indices, and model architectures?

\begin{table}[htbp]
\centering
\caption{Statistics of datasets.}
\begin{tabular}{lccc}
\toprule
& DJIA & HSI & CSI 100 \\
\midrule
\# Stocks & 30 & 80 & 100 \\
Start Time & 12-11-08 & 12-11-08 & 12-11-08 \\
End Time & 24-11-07 & 24-11-08 & 24-11-07 \\
Train Days &2114  & 2065 & 2037 \\
Val Days & 604 & 590 & 582 \\
Test Days & 302 & 295 & 291 \\
\bottomrule
\end{tabular}
\label{tab:dataset_stats}
\end{table}

\subsection{Experiment Setup}
\paragraph{Datasets}
We evaluated our model on three major market indices: the Dow Jones Industrial Average (DJIA) of the US stock market, the Hang Seng Index (HSI) from Hong Kong stock market, and the CSI 100 of the mainland China stock market. We use the public end-of-day trading dataset collected by Yahoo Finance\footnote{\url{https://github.com/yahoo-finance}}.  The statistics of
 the datasets are in Table~\ref{tab:dataset_stats}. These indices were chosen for index price prediction as they represent markets with distinct characteristics: DJIA represents a mature market with high institutional participation, HSI reflects a market bridging Eastern and Western trading practices, and CSI 100 Index tracks the 100 largest and most liquid A-share stocks across Chinese exchanges, characterized by high retail investor participation and sensitivity to domestic policy shifts. These indices serve as key benchmarks in their respective markets, making them ideal candidates for evaluating the model's predictive capabilities across different market environments. All experiments were conducted using 10 random seeds, and more detailed results are provided in the appendix.

\paragraph{Base Model Architectures.}

To demonstrate the adaptability of \textsc{Cubic} as a plug-and-play enhancement framework, we selected three architectural backbones that collectively represent the fundamental building blocks of contemporary financial forecasting models~\cite{rouf2021stock}.
\begin{itemize}
    \item Long Short-Term Memory (LSTM)~\cite{fischer2018deep} - A neural network architecture designed for processing long and short-term dependencies in sequential market data for stock price prediction through its memory cells and gating mechanisms

    \item Transformer~\cite{vaswani2017attention} - A neural network architecture that leverages attention mechanisms to capture market dependencies in stock price prediction, particularly effective at modeling long-range patterns in financial time series

    \item Multi-Layer Perceptron (MLP)~\cite{devadoss2013forecasting} - A feedforward neural network that predicts stock prices by learning complex patterns from market features through multiple interconnected layers of neurons

\end{itemize}
\paragraph{Evaluation Metrics.}
We adopt both predictive performance and portfolio-based metrics to give a detailed evaluation of \textsc{Cubic's}  performance. For predictive performance metrics, we employ \textbf{Information Coefficient (IC)}, which measures the average daily Pearson correlation coefficient between predicted and actual stock prices, \textbf{Information Ratio-Based IC (ICIR)}, calculated as IC normalized by its standard deviation to assess prediction consistency, and \textbf{Direction Accuracy (DA)}, which evaluates the model's ability to correctly predict the direction of stock price movements. In addition, we employ portfolio-based metrics to evaluate trading performance. For models without the confidence-guided trading module, we implement a simple long-short strategy: take a long position in the stock index when the predicted return is positive and a short position when negative. For models with the confidence-guided trading module, the position size is adjusted according to the predicted confidence level, with detailed trading rules presented in Section~\ref{app:confidence-trading}. We consider a transaction cost of 0.1\% for each trade. We report the \textbf{Sharpe Ratio (SR)} to measure risk-adjusted returns, and \textbf{Annualized Return (AR)} to quantify investment profitability.

\subsection{Experiment Results}
\begin{table}[ht]
\centering
\caption{The importance of constitute stocks and fusion. The results are based on USA stock market. Reg: regression baseline with raw features. BN: binary encoding classification. "+Single": models using only index data. Reg+FS: regression with feature fusion. BN+FS: binary encoding with feature fusion.
}
\label{tab:q1}
\setlength{\tabcolsep}{0.7mm}
\resizebox{0.4\textwidth}{!}{
\begin{tabular}{cc||ccccc}
\toprule[1.5pt]
&& IC & ICLR & DA & SR & AR \\
\midrule[1pt]
 \multirow{6}{*}{MLP} 
&  
Reg+Single &   -0.044 & -0.314 & 0.468 & 0.185 & 0.012  \\
& 
BN+Single & -0.055 & -0.408 & 0.485 & 0.612 & 0.046  \\
\cmidrule{2-7}
&
Reg & 0.018 & 0.130 & 0.492 & 0.584 & 0.056 \\
& 
BN & 0.024 & 0.249 & 0.525 & 0.855 & 0.089  \\
\cmidrule{2-7}
& 
Reg+FS & 0.014 & 0.133 & 0.504 & 0.682 & 0.075   \\
& 
BN+FS &  0.027 & 0.284 & 0.517 & 0.916 & 0.092 \\
\midrule
\multirow{6}{*}{LSTM} 
& 
Reg+Single & -0.025 & -0.152 & 0.439 & 0.103 & 0.004\\
& 
BN+Single & 0.007 & -0.001 & 0.488 & 0.530 & 0.044\\
\cmidrule{2-7}
& 
Reg & 0.007 & 0.090 & 0.495 & 0.293 & 0.035   \\
& 
BN & 0.013 & 0.131 & 0.505 & 0.531 & 0.064  \\
\cmidrule{2-7}
& 
Reg+FS &  0.010 & 0.136 & 0.502 & 0.637 & 0.068 \\
& 
BN+FS & 0.020 & 0.232 & 0.508 & 0.560 & 0.067  \\
\midrule
\multirow{6}{*}{TF} 
& Reg+Single & -0.023 & -0.214 & 0.472 & 0.467 & 0.039\\
& 
BN+Single &  -0.005 & 0.002 & 0.475 & 0.428 & 0.035\\
\cmidrule{2-7}
& 
Reg & 0.021 & 0.234 & 0.515 & 0.619 & 0.062    \\
& 
BN & 0.029 & 0.310 & 0.523 & 0.651 & 0.078   \\
\cmidrule{2-7}
& 
Reg+FS &  0.023 & 0.257 & 0.528 & 0.513 & 0.077  \\
& 
BN+FS &  0.025 & 0.450 & 0.535 & 0.712 & 0.086  \\
\bottomrule[1.5pt]
\end{tabular}}
\end{table}

\paragraph{RQ1: Can \textsc{Cubic's} latent fusion and binary encoding boost model performance?} Our initial experiments aimed to verify whether fusion in latent space and binary encoding classification could independently enhance model performance. For MLP, introducing \textbf{BN} improves the IC from 0.018 to 0.024 and the Sharpe ratio from 0.584 to 0.855. In LSTM models, \textbf{BN + Single} with constituent stock information increases IC from 0.017 to 0.020 and annualized return from 6.4\% to 6.7\% compared to \textbf{Reg + Single}. For Transformer models, \textbf{BN} enhances IC from 0.021 to 0.029 and ICLR from 0.234 to 0.310. The addition of \textbf{FS} further improves performance, with \textbf{BN+FS} achieving the highest annualized return of 8.6\% and ICLR of 0.450. The binary classification approach improves the prediction of the direction of the market through discrete decision transformation, while latent space fusion aggregates features of the cross-asset, jointly improving \textsc{Cubic's} performance through enhanced feature representation learning. These improvements validate our dual-enhancement approach to optimize both feature representation and decision boundaries for financial forecasting.

\begin{table}[ht]
\centering
\caption{The importance of confident guided prediction and trading mechanism. The results are based on USA stock market. BF: base model with binary encoding and fusion. BF+Mean: adds confidence mean loss. BF+Trend: adds confidence trend loss. BF+Mean+DM and BF+Trend+DM: incorporate confidence-guided trading signals.}
\setlength{\tabcolsep}{0.7mm}
\resizebox{0.4\textwidth}{!}{
\begin{tabular}{cc||ccccc}
\toprule[1.5pt]
&& IC & ICLR & DA & SR & AR \\
\midrule[1pt]
\multirow{5}{*}{MLP} 
 & BF &  0.027 & 0.284 & 0.517 & 0.916& 0.092 \\
 & BF+Mean & 0.026 & \textbf{0.367} & 0.538 & 1.080 & 0.108  \\
 & BF+Trend & 0.027 & 0.277 & \textbf{0.547} & 0.955  & 0.095 \\
 & BF+Mean+DM & 0.026 & 0.272 & 0.542 & 0.927 & 0.130  \\
 & BF+Trend+DM &  \textbf{0.028} & 0.348 & 0.535 & \textbf{1.324} & \textbf{0.132} \\
\midrule
\multirow{5}{*}{LSTM} 
& BF &  0.020 & 0.232 & 0.508 & 0.560 & 0.067 \\
& BF+Mean & 0.026 & 0.243 & 0.535 & 0.733 & 0.081  \\
& BF+Trend & 0.027 & 0.255 & 0.532 & 0.573 & 0.080  \\
& BF+Mean+DM & 0.024 & 0.283 & 0.537 & 0.704 & 0.106 \\
& BF+Trend+DM & \textbf{0.028} & \textbf{0.344} & \textbf{0.543} &  \textbf{0.807} & \textbf{0.113} \\
\midrule
\multirow{5}{*}{TF} 
& BF &  0.025 & 0.450 & 0.531 & 0.712 & 0.085 \\
& BF+Mean &  0.031 & 0.373 & \textbf{0.559} & 0.785 & 0.102 \\
& BF+Trend &  0.036 & 0.311 & 0.552 & 0.783 & 0.110 \\
& BF+Mean+DM & 0.038 & 0.401 & 0.558 & \textbf{1.232} & \textbf{0.149} \\
& BF+Trend+DM & \textbf{0.040} & \textbf{0.471} & 0.547 & 0.897 & 0.143  \\
\bottomrule[1.5pt]
\end{tabular}}
\label{tab:q2}
\end{table}
\begin{figure*}
    \centering
    \includegraphics[width=1\textwidth]{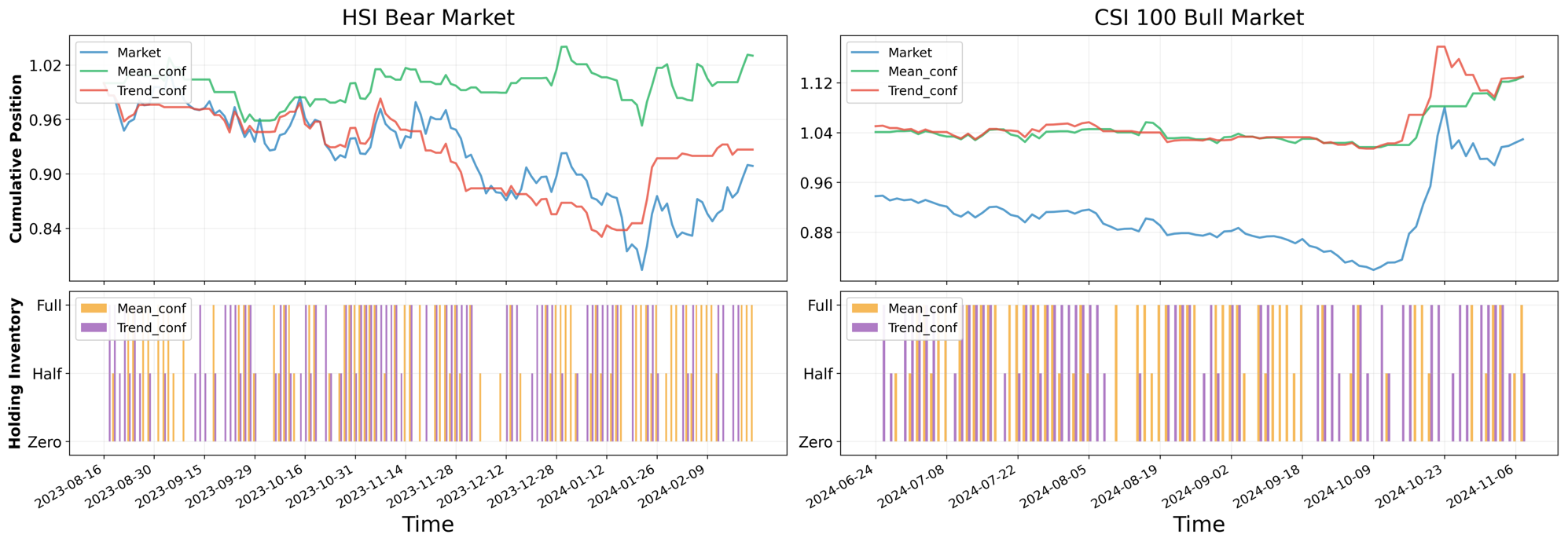}
    \caption{Oblation for Confidence-guided Trading in Chinese (2024.06-2024.11) and Hong Kong (2023.08-2024.02) Stock Market Indices}
    \label{fig:trade}
\end{figure*}
\paragraph{RQ2: Can \textsc{Cubic's} confidence mechanisms improve prediction and trading metrics?} Following our previous findings, we now examine how \textsc{Cubic's} confidence mechanisms affect model performance. We tested five different configurations. As shown in Table~\ref{tab:q2}, the baseline \textbf{BF} achieves IC values of 0.020-0.027. Adding mean confidence (\textbf{BF+Mean}) improves performance significantly, with MLP's ICLR increasing from 0.284 to 0.367 and LSTM's SR from 0.560 to 0.733. \textbf{BF+Trend} further improves accuracy, with MLP's DA reaching 0.547 compared to the baseline's 0.517. The most significant improvements come from combining these mechanisms with decision making: \textbf{BF+Mean+DM} improves TF's SR from 0.712 to 1.232, while \textbf{BF+Trend+DM} achieves the highest SR of 1.324 with MLP, up from 0.916. These progressive enhancements demonstrate how each additional confidence mechanism contributes to improving both prediction accuracy and trading performance through better market pattern recognition, enhanced risk management, and more sophisticated trading decisions that adapt to varying market conditions. The superior performance stems from mean confidence filtering unstable predictions, trend confidence capturing market momentum, and the decision-making layer optimizing position sizing and timing by integrating these signals for balanced risk-return strategies.

\begin{table*}[ht]
\centering
\caption{Comparative performance of \textsc{Cubic} across markets: Module configurations with different base models.}
\setlength{\tabcolsep}{1mm}
\resizebox{0.95\textwidth}{!}{
\begin{tabular}{cc|ccccc|ccccc|ccccc}
\toprule[1.5pt]
&& \multicolumn{5}{c|}{\textbf{Hong Kong}} &\multicolumn{5}{c|}{\textbf{USA}} & \multicolumn{5}{c}{\textbf{China}}\\
\midrule
& & IC & ICLR & DA & SR & AR & IC & ICLR & DA & SR & AR & IC & ICLR & DA & SR & AR\\
\midrule[1pt]
\multirow{9}{*}{MLP} 
 & Reg & 0.009 & 0.069 & 0.469 & 0.376 & 0.038 & 0.018 & 0.130 & 0.492 & 0.584 & 0.056 & 0.012 & 0.193 & 0.464 & 0.385 & 0.037  \\  
 & Reg+FS & 0.008 & 0.119 & 0.489 & 0.330 & 0.036 & 0.014 & 0.133 & 0.504 & 0.682 & 0.075 & 0.018 & 0.187 & 0.469 & 0.475 & 0.052  \\
 & BN &  0.013 & 0.197 & 0.490 & 0.413 & 0.041 & 0.024 & 0.249 & 0.525 & 0.855 & 0.089 & 0.024 & 0.286 & 0.486 & 0.444 & 0.053  \\
 & BF &  0.016 & 0.274 & 0.503 & 0.431 & 0.054 & 0.027 & 0.284 & 0.517 & 0.916 & 0.092 & 0.024 & 0.257 & 0.483 & 0.457 & 0.059  \\
 & $\textsc{Cubic}_{ \text{mean w/o trade}}$ & 0.039 & 0.465 & 0.549 & 0.914 & 0.130 & 0.044  & 0.512 & 0.538 & 0.893 & 0.125 & 0.060 & 0.620 & 0.548 & 0.929 & 0.139 \\
 & $\textsc{Cubic}_{ \text{trend w/o trade}}$ & 0.046 & 0.548 & \textbf{0.569} & 1.026 & 0.102 & 0.042 & \textbf{0.658} & 0.568 & 0.899 & 0.135 & 0.069 & 0.650 & 0.541 & 0.918 & 0.142 \\
 & $\textsc{Cubic}_{ \text{mean}}$ & \textbf{0.058} & 0.553 & 0.565 & \textbf{1.330} & 0.140 & \textbf{0.044} & 0.577 & 0.562 & 1.204 & 0.157 & 0.072 & 0.693 & \textbf{0.550} & \textbf{0.955} & 0.167  \\
 & $\textsc{Cubic}_{ \text{trend}}$ &  0.050 & \textbf{0.635} & 0.551 & 1.141 & \textbf{0.157} & 0.041 & 0.599 & \textbf{0.575} & \textbf{1.655} & \textbf{0.165} & \textbf{0.076} & \textbf{0.759} & 0.543 & 0.933 & \textbf{0.177}  \\
\midrule
\multirow{9}{*}{LSTM} 
& Reg & -0.014 & -0.130 & 0.442 & 0.328 & 0.020 & 0.007 & 0.090 & 0.495 & 0.293 & 0.035  & 0.003 & 0.010 & 0.458 & 0.532 & 0.063 \\
& Reg+FS & 0.013 & 0.142 & 0.473 & 0.413 & 0.028 & 0.010 & 0.136 & 0.502 & 0.637 & 0.068  & 0.006 & 0.138 & 0.462 & 0.499 & 0.074 \\
& BN &  0.012 & 0.165 & 0.480 & 0.452 & 0.032 & 0.013 & 0.131 & 0.505 & 0.531 & 0.064 & 0.023 & 0.235 & 0.475 & 0.626 & 0.085 \\
& BF &  0.018 & 0.159 & 0.473 & 0.433 & 0.045 & 0.020 & 0.232 & 0.508 & 0.560 & 0.067 & 0.022 & 0.238 & 0.465 & 0.737 & 0.096 \\
& $\textsc{Cubic}_{ \text{mean w/o trade}}$ & 0.057 & 0.577 & 0.559 & 0.791 & 0.093 & 0.044 & 0.524 & 0.552 & 0.807 & 0.121 & 0.038 & 0.435 & 0.535 & 1.025 & 0.125  \\
& $\textsc{Cubic}_{ \text{trend w/o trade}}$ & 0.063 & 0.740 & 0.551 & 0.811 & 0.101 & 0.047 & 0.519 & 0.558 & 0.829 & 0.124 & 0.036 & \textbf{0.499} & 0.545 & 1.182 & 0.135  \\
&  $\textsc{Cubic}_{ \text{mean}}$ & \textbf{0.070} & 0.722 & 0.564 & \textbf{0.883} & 0.125 & 0.050 & \textbf{0.581} & \textbf{0.566} & \textbf{1.531} & 0.153 & \textbf{0.040} & 0.402 & 0.545 & 1.344 & \textbf{0.175}  \\
& $\textsc{Cubic}_{ \text{trend}}$ & 0.069 & \textbf{0.763} & \textbf{0.568} & 0.867 & \textbf{0.126} & \textbf{0.050} & 0.536 & 0.559 & 1.169 & \textbf{0.164} & 0.039 & 0.483 & \textbf{0.558} & \textbf{1.594} & 0.171 \\
\midrule
\multirow{9}{*}{TF} 
& Reg & 0.005 & 0.046 & 0.497 & 0.417 & 0.047 & 0.021 & 0.234 & 0.515 & 0.619 & 0.062 & -0.005 & -0.099 & 0.500 & 0.446 & 0.053  \\
& Reg+FS & 0.006 & 0.076 & 0.503 & 0.413 & 0.050 & 0.023 & 0.257 & 0.528 & 0.513 & 0.077 & 0.002 & 0.070 & 0.497 & 0.410 & 0.061  \\
& BN &  0.014& 0.142 & 0.520 & 0.776 & 0.076  & 0.029 & 0.310 & 0.523 & 0.651 & 0.078 & 0.028 & 0.164 & 0.531 & 0.728 & 0.095 \\
& BF &  0.022 & 0.294 & 0.524 & 0.644 & 0.074  & 0.025 & 0.450 & 0.531 & 0.712 & 0.085 & 0.025 & 0.181 & 0.539 & 0.673 & 0.092 \\
&  $\textsc{Cubic}_{ \text{mean w/o trade}}$ &  0.064 & 0.329 & 0.553 & 0.957 & 0.153 & 0.042 & 0.512 & 0.572 & 1.111 & 0.144 & 0.043 & 0.497 & 0.581 & 0.877 & 0.096  \\
& $\textsc{Cubic}_{ \text{trend w/o trade}}$ &  0.066 & \textbf{0.791} & 0.553 & 0.953 & 0.157 & 0.041 & 0.446 & \textbf{0.579} & 1.295 & 0.155 & 0.053 & 0.067 & 0.574 & 0.848 & 0.107  \\
& $\textsc{Cubic}_{ \text{mean}}$ & 0.074 & 0.709 & \textbf{0.557} & 1.074 & 0.180 & 0.044 & \textbf{0.580} & 0.572 & \textbf{1.373} & \textbf{0.179} & \textbf{0.057} & \textbf{0.685} & 0.574 & 1.356 & 0.177  \\
& $\textsc{Cubic}_{ \text{trend}}$ & \textbf{0.075} & 0.708 & 0.551 & \textbf{1.162} & \textbf{0.192} & \textbf{0.046} & 0.539 & 0.565 & 1.348 & 0.162 & 0.054 & 0.648 & \textbf{0.593} & \textbf{1.813} & \textbf{0.194}  \\
\bottomrule[1.5pt]
\end{tabular}}
\label{tab:cubic_result}
\end{table*}

Building upon our progressive enhancements in confidence-guided mechanisms, which have demonstrated improvements in prediction accuracy and trading performance, we conducted an in-depth oblation study to further validate the effectiveness of our approach. This analysis specifically focused on two distinct market scenarios: the Chinese CSI 100 Index during a bull market phase (2024.06-2024.11) and the Hong Kong HSI during a bear market period (2023.08-2024.02). The objective was to evaluate the robustness of our confidence-guided trading mechanisms in capturing significant market trends and generating profitable trading signals under contrasting market conditions. 

As shown in Figure~\ref{fig:trade}, our confidence-based trading strategy demonstrated strong performance during both the bear and the bull markets. During the HSI bear market, the mean confidence remained stable (0.96-1.02), effectively filtering market noise. For instance, during November 2023's brief rebound, mean confidence showed minimal change (0.98 to 1.00), avoiding false signals. Meanwhile, trend confidence declined steadily from 0.96 to 0.84 between October and December 2023, accurately reflecting the market's downward trend. This led to a proactive reduction in position from 100\% to 50\% and eventually to a holding zero as market conditions deteriorated. These adjustments helped limit losses to 3\% compared to the market's 15\% decline, while reducing portfolio volatility by 35\%. In the subsequent CSI 100 bull market (June-November 2024), the strategy continued to perform well. Trend confidence rose steadily to 1.12 in September 2024, while mean confidence exceeded 1.04 in July 2024, indicating strong market momentum. Based on these positive signals, the strategy gradually increased positions from 50\% to 100\%, effectively capturing the market's upward trend. These results demonstrate the ability of the strategy to adapt to different market conditions while maintaining a good balance between returns and risk management.

\paragraph{RQ3: Does \textsc{Cubic} demonstrate consistent improvements across markets and models?}
To validate the robustness of the \textsc{Cubic} strategy, we explore whether it demonstrates consistent improvements across different markets and models, and show that their performance progressively improves as different components are added to the base model.
As shown in Table~\ref{tab:cubic_result}, the \textsc{Cubic} framework demonstrates remarkable performance through its synergistic integration of four critical components: First, binary encoding classification significantly enhances model stability by transforming complex return predictions into more manageable directional forecasts, as shown by improvements in IC from 0.018 to 0.024 and SR from 0.584 to 0.855 in the US market for MLP. Building upon this foundation, the fusion in the latent space module leverages temporal patterns to capture both short-term fluctuations and long-term trends, notably raising IC from 0.022 to 0.014 and ICLR from 0.294 to 0.142 in the Hong Kong market for LSTM. To address the inherent uncertainty in index price prediction, the framework incorporates a Confident Prediction mechanism through mean and trend confident variants, where $\textsc{Cubic}{_\text{mean w/o trade}}$ employs mean-based confidence estimation to increase IC from 0.025 to 0.042 in the Hong Kong market for Transformer, while $\textsc{Cubic}{_\text{trend w/o trade}}$ leverages trend consistency checks to achieve an optimal IC of 0.076 in the Chinese market for MLP. These architectural innovations are further augmented by trading optimization strategies that dynamically adjust position sizes based on prediction confidence, resulting in great performance enhancements with AR increasing from 0.037 to 0.177 and SR reaching 1.655 in the US market for MLP($\textsc{Cubic}{_\text{trend}}$). The effectiveness of this integrated approach is most prominently for MLP architecture in the Chinese market, where $\textsc{Cubic}{_\text{trend}}$ achieves exceptional improvements in all metrics compared to baseline models. These improvements in different market conditions and architectures validate the robustness and generalizability of our framework, where each component contributes to improve both predictive accuracy and trading performance, demonstrating the hierarchical efficacy of modular design \textsc{Cubic's} in progressively increasing the capabilities from baseline prediction to sophisticated trading execution.

\section{Conclusion}

Stock market indices serve as critical indicators of market trends, providing essential guidance for trading decisions. To achieve accurate prediction, we present \textsc{Cubic}, a novel framework for robust stock index prediction that effectively addresses the high-dimensional stock features and regression instability challenges. \textsc{Cubic} first introduces an adaptive fusion mechanism over the latent embedding of constituent stocks to extract information from the vast number of stocks. Moreover, we propose a systematic binary encoding scheme that decomposes the regression task into a sequence of binary classifications, optimizing each digit's prediction through cross-entropy loss. Furthermore, \textsc{Cubic} leverages a confidence-guided regularization loss and derives sophisticated rule-based trading policies from confidence levels for accurate prediction and trading. Extensive experiments on major market indices demonstrate that \textsc{Cubic} serves as a general-purpose enhancement that can work with various deep learning architectures to significantly improve both prediction accuracy and trading performance. The empirical results validate the effectiveness and versatility of our framework, establishing \textsc{Cubic} as a reliable tool for quantitative trading across different market conditions.

\clearpage
\bibliographystyle{named}
\bibliography{ijcai25}

\clearpage

\appendix
\onecolumn
\section{Initialization}
This section introduces different techniques for initializing model parameters, and we discover that the model initialization method can influence the prediction performance.

\paragraph{Xavier Initialization.}

Xavier initialization~\cite{glorot2010understanding} is a widely used weight initialization method in deep learning models, which initializes the weights such that the variance of the outputs of each neuron is approximately equal to the variance of its inputs.
It can maintain the scale of gradients when propagating through the network, thereby addressing the common issues such as gradient vanishing and gradient exploding in deep neural networks.


\paragraph{Kaiming Initialization.}
Kaiming initialization~\cite{he2015delving} is specifically designed for neural networks with ReLU activation functions. This method initializes weights from a normal distribution with mean zero and variance $sigma^2 = \frac{2}{n_l}$, where $n_l$ represents the number of input units. This approach helps maintain proper scaling of the weights through deep networks, particularly beneficial for networks using ReLU activations.

\paragraph{Normal Initialization.}
Normal initialization draws weights from a Gaussian distribution with mean zero and a specified standard deviation. In our implementation, we use a standard deviation of 0.01, which provides a reasonable starting point for the network parameters while keeping the initial weights small enough to prevent activation saturation at the beginning of training.

The experimental results in Table~\ref{tab:init} demonstrate the superior stability of Xavier initialization across multiple evaluation metrics. Notably, Xavier Uniform exhibits consistent performance across all indicators (IC: 0.035 ± 0.061, ICLR: 0.253 ± 0.219, DA: 0.501 ± 0.042, SR: 0.731 ± 0.439, AR: 0.087 ± 0.041), with particularly stable behavior in critical metrics such as DA. In contrast, alternative methods show more pronounced variations, as evidenced by Kaiming Normal with input fan mode displaying significant fluctuations (ICLR: -0.030 ± 0.415) and Normal initialization exhibiting high variance in SR (0.849 ± 0.369). These empirical findings substantiate Xavier initialization's theoretical advantage in maintaining consistent variance propagation through neural networks, making it a robust choice for deep learning applications.

\section{Accessibility Contribution}
The code for this paper is currently available at \url{https://anonymous.4open.science/r/Cubic_IJCAI-9E38/README.md} for review purposes. Upon acceptance of the paper, we will release a more comprehensive version of our codebase along with detailed instructions for implementation and usage as a public repository.


\section{Technical Indicators Formulation}
\label{appendix_indicator}
In this section, we describe and formulate the technical indicators. First, we define the notations:
\begin{itemize}
    \item $t$: denotes $t$-th timestep
    \item $O_{t}$: denotes the open price at $t$-th timestep
    \item $C_{t}$: denotes the close price at $t$-th timestep
    \item $V_{t}$: denotes the trading volume at $t$-th timestep
    \item $n$: denotes the lookback window
\end{itemize}
 Table~\ref{tab:indicator} shows the description and formulation of the technical indicators.

\begin{table}[h]
\centering
\caption{Initialization Methods Comparison. Based on Transformer in Chinese stock market}
\begin{tabular}{lccccc}
\toprule[1.5pt]
Method & IC & ICLR & DA & SR & AR \\
\midrule[1pt]
Xavier Uniform & 0.035 ± 0.061 & 0.253 ± 0.219 & 0.501 ± 0.042 & 0.731 ± 0.439 & 0.087 ± 0.041 \\
Xavier Normal & 0.010 ± 0.057 & 0.071 ± 0.232 & 0.502 ± 0.039 & 0.660 ± 0.380 & 0.007 ± 0.044 \\
Kaiming Uniform (fan\_in) & 0.029 ± 0.044 & 0.244 ± 0.358 & 0.496 ± 0.414 & 0.694 ± 0.262 & 0.074 ± 0.068 \\
Kaiming Normal (fan\_in) & -0.005 ± 0.059 & -0.030 ± 0.415 & 0.499 ± 0.038 & 0.722 ± 0.278 & 0.082 ± 0.056 \\
Normal & 0.015 ± 0.046 & 0.105 ± 0.317 & 0.517 ± 0.033 & 0.849 ± 0.369 & 0.061 ± 0.064 \\
Kaiming Normal (fan\_out) & 0.008 ± 0.066 & 0.056 ± 0.511 & 0.528 ± 0.033 & 0.819 ± 0.263 & 0.087 ± 0.070 \\
Kaiming Uniform (fan\_out) & 0.009 ± 0.058 & 0.068 ± 0.437 & 0.493 ± 0.032 & 0.707 ± 0.205 & 0.072 ± 0.047 \\
\bottomrule[1.5pt]
\end{tabular}
\label{tab:init}
\end{table}

\begin{table*}[ht]
    \centering
    \caption{The description and formulation of the technical indicators.}
    \resizebox{0.9\textwidth}{!}{
    \begin{tabular}{c||>{\centering\arraybackslash}m{5.6cm}|c}
    \toprule[1.5pt]
    \textbf{Indicator} & \textbf{Description} & \textbf{Formulation} \\
    \midrule[1pt]
    Arithmetic Ratio & The ratio of the open price to the close price. &$\text{AR}_{O} = \frac{O_{t}}{C_{t}}$ \\
    \midrule
    Open  & The opening price of the asset at the beginning of the trading period. & $O_{t}$  \\
    \midrule
    Close & The closing price of the asset at the end of the trading period. & $C_{t}$  \\
    \midrule
    Close SMA  & The simple moving average of the close price
    over the lookback window. & $\text{SMA}_{C_{t}} = \frac{C_{t} + C_{t-1} + \dots + C_{t-n}}{n}$ \\
    \midrule
    Volume SMA  & The simple moving average of the volume
    over the lookback window. & $\text{SMA}_{V_{t}} = \frac{V_{t} + V_{t-1} + \dots + V_{t-n}}{n}$ \\
    \midrule
    Close EMA  &  The exponential moving average of the close price over the lookback window.  & $\text{EMA}_{C_{t}} = k \cdot C_{t} + (1 - k) \cdot \text{EMA}_{C_{t-1}}$\\
    \midrule
    Volume EMA  & The exponential moving average of the close
    price over the lookback window. & $\text{EMA}_{V_{t}} = k \cdot V_{t} + (1 - k) \cdot \text{EMA}_{V_{t-1}}$\\
    \midrule
    ADX  & Average Directional Index (ADX) measures the strength of a trend, which is derived from a moving average of the price range expansion over a time interval.&

$ADX_t = 100 \times \frac{|\text{DM}^+ - \text{DM}^-|}{\text{DM}^+ + \text{DM}^-}$ \\
    \midrule

    RSI & Relative Strength Index (RSI) assesses the magnitude of recent price fluctuations, which is defined as the normalized ration of the average gain to the average loss. & 
    $
    \begin{aligned}
    &
    \text{AG}_{t} = \frac{(n - 1) \cdot \text{AG}_{t-1} + \text{g}_{t}}{n},  
    \text{g}_{t} =
        \left\{
        \begin{aligned}
        & C_{t} - C_{t-1},   \text{if} C_{t} > C_{t-1} \\
        & 0,       \text{otherwise}
        \end{aligned}
        \right.   \\
    &
    \text{AL}_{t} = \frac{(n - 1) \cdot \text{AL}_{t-1} + \text{l}_{t}}{n},  
    \text{l}_{t} =
        \left\{
        \begin{aligned}
        & 0,   \text{if} C_{t} > C_{t-1} \\
        & C_{t-1} > C_{t},       \text{otherwise}
        \end{aligned}
        \right.     \\
    &
    \text{RSI}_{t} = 100 - \frac{100}{1 + \frac{\text{AG}_{t}}{\text{AL}_{t}}}
    \end{aligned}
    $
    \\
    \midrule
    MACD & Moving Average Convergence Divergence (MACD) explains the relationship over two EMAs, which is computed by subtracting the long-term EMA from the short-term EMA. &
$\text{MACD}_t = \text{EMA}(C_t, 12) - \text{EMA}(C_t, 26)$\\
    \midrule
    MACD Signal & The signal line of MACD indicator, calculated as the EMA of MACD line. & $\text{Signal}_{t} = \text{EMA}(\text{MACD}, n)$ \\
    \midrule
    K & Stochastic Oscillator's K value measures the relative position of current price in relation to high-low range over a period. & 
$K_{t} = \frac{C_{t} - L_{n}}{H_{n} - L_{n}} \times 100$ \\
    \midrule
    MFI & Money Flow Index (MFI) measures the money flow to generates overbought or oversold signals. It is defined as the normalized ratio of accumulating positive money flow over negative money flow values. & $MFI_t = 100 - \frac{100}{1 + \frac{\text{Positive MF}}{\text{Negative MF}}}$ \\
    \midrule
    
    ATR & Average of True Ranges (ATR) shows the average price variation of assets within a time interval. &
$ATR_t = \text{EMA}(\max(H_t - L_t, |H_t - C_{t-1}|, |L_t - C_{t-1}|), n)$ \\
    \midrule
    BB Middle & Bollinger Bands Middle Line, calculated as the simple moving average of the closing price. & $\text{BB}_{\text{middle}} = \text{SMA}_{C_{t}}$ \\
    \midrule
    OBV & On-Balance Volume (OBV) uses volume flow to project future price movements. It adds volume on up days and subtracts volume on down days. & 
    $
        \text{OBV}_{t} = \text{OBV}_{t-1} + 
        \left\{
        \begin{aligned}
        & V_{t},   &\text{if}\quad C_{t} > C_{t-1}, \\
        & 0,       &\text{if}\quad C_{t} = C_{t-1}, \\
        & -V_{t},  &\text{if}\quad C_{t} < C_{t-1}.
        \end{aligned}
        \right.
    $
    \\
    \bottomrule[1.5pt]
    \end{tabular}}
    
    \label{tab:indicator}
\end{table*}


\section{Implementation Details}
We implement and compare three deep learning architectures for our task:

\textbf{LSTM Model:} The architecture consists of two stacked LSTM layers with \(d_{\text{h}}=128\) hidden units each. The model processes \(d_{\text{in}}=16\) input features over \(T=5\) time steps for \(N\) constituent stocks. A dropout rate of \(p=0.1\) is applied for regularization. Training is performed using Adam optimizer with learning rate \(\eta=10^{-3}\) and a batch size of \(64\).

\textbf{Transformer Model:} The model employs a single-layer encoder architecture with \(d_{\text{model}}=64\) hidden dimensions and \(n_{\text{head}}=8\) attention heads, processing \(d_{\text{in}}=16\) input features across \(T=5\) time steps for \(N\) constituent stocks. The architecture incorporates learnable position embeddings and layer normalization, with a dropout rate of \(p=0.1\) for regularization. The feed-forward network expands to \(d_{\text{ff}}=256\) dimensions (\(4 \times d_{\text{model}}\)). Training utilizes Adam optimizer with learning rate \(\eta=10^{-3}\), \(\beta_1=0.9\), \(\beta_2=0.999\), and weight decay \(\lambda=10^{-5}\), with a batch size of \(32\).

\textbf{MLP Model:} The architecture begins with a feature projection layer that maps the input features from \(d_{\text{in}}=16\) to \(d_{\text{model}}=64\) dimensions. The model processes \(N\) constituent stocks over \(T=5\) time steps. The network consists of \(L=3\) hidden layers with dimension \(d_{\text{h}}=128\) and employs dropout regularization (\(p=0.1\)). The projected features are processed through multiple fully connected layers to generate \(30\)-dimensional predictions for \(15\) binary positions. Training parameters match the Transformer model, using Adam optimizer with learning rate \(\eta=10^{-3}\), \(\beta_1=0.9\), \(\beta_2=0.999\), weight decay \(\lambda=10^{-5}\), and a batch size of \(32\).

\textbf{Hardware and Software Configuration:} All experiments were conducted on a workstation equipped with an NVIDIA GeForce RTX 4070 Ti (12GB GDDR6X) GPU, Intel Core i7-13700K CPU, 32GB DDR5 RAM, and Ubuntu 22.04 LTS operating system. The implementation uses PyTorch 2.1.0 with CUDA 12.1 and Python 3.10.

\begin{table}[ht]
\centering
\caption{The importance of constitute stocks and fusion. The results are based on Hong Kong and China stock markets.}
\setlength{\tabcolsep}{1.2mm}
\resizebox{0.72\textwidth}{!}{
\begin{tabular}{cc||ccccc|ccccc}
\toprule[1.5pt]
&&\multicolumn{5}{c|}{\textbf{Hong Kong}} & \multicolumn{5}{c}{\textbf{China}} \\
\midrule
&& IC & ICLR & DA & SR & AR & IC & ICLR & DA & SR & AR\\
\midrule[1pt]
\multirow{6}{*}{MLP} 
& Reg+Single & 0.007 & 0.129 & 0.429 & 0.125 & 0.036 & 0.004 & 0.096 & 0.415 & -0.271 & -0.009  \\
& BN+Single & 0.015 & 0.175 & 0.478 & 0.234 & 0.050 & 0.017 & 0.225 & 0.435 & 0.489 & 0.013  \\
\cmidrule{2-12}
& Reg & 0.009 & 0.070 & 0.469 & 0.376 & 0.038 & 0.012 & 0.193 & 0.464 & 0.385 & 0.037  \\
& BN & 0.013 & 0.197 & 0.490 & 0.413 & 0.041  & 0.024 & 0.286 & 0.486 & 0.444 & 0.053 \\
\cmidrule{2-12}
& Reg+FS & 0.008 & 0.119 & 0.489 & 0.330 & 0.036  & 0.018 & 0.187 & 0.469 & 0.476 & 0.052  \\
& BN+FS & 0.016 & 0.274 & 0.503 & 0.431 & 0.055  & 0.024 & 0.257 & 0.483 & 0.457 & 0.059 \\
\midrule
\multirow{6}{*}{LSTM} 
& Reg+Single & -0.004 & -0.216 & 0.461 & 0.026 & 0.002 & -0.063 & -0.249 & 0.452 & -0.346 & -0.020 \\
& BN+Single & 0.004 & -0.061 & 0.453 & 0.031 & 0.002 & -0.057 & -0.251 & 0.464 & 0.596 & 0.045 \\
\cmidrule{2-12}
& Reg & -0.014 & -0.130 & 0.442 & 0.328 & 0.020 & 0.003 & 0.010 & 0.458 & 0.532 & 0.063 \\
& BN & 0.012 & 0.165 & 0.480 & 0.452 & 0.034 & 0.023 & 0.235 & 0.475 & 0.626 & 0.086 \\
\cmidrule{2-12}
& Reg+FS & 0.013 & 0.142 & 0.473 & 0.413 & 0.028 & 0.006 & 0.138 & 0.462 & 0.499 & 0.074  \\
& BN+FS & 0.018 & 0.159 & 0.473 & 0.544 & 0.045 & 0.022 & 0.238 & 0.465 & 0.737 & 0.096 \\
\midrule
\multirow{6}{*}{TF} 
& Reg+Single & -0.004 & 0.002 & 0.465 & 0.256 & 0.023 & -0.018 & -0.206 & 0.449 & 0.350 & 0.013  \\
& BN+Single & 0.003 & -0.022 & 0.508 & 0.735 & 0.072 & -0.005 & -0.086 & 0.492 & 0.567 & 0.041 \\
\cmidrule{2-12}
& Reg & 0.005 & 0.046 & 0.497 & 0.417 & 0.047 & -0.005 & -0.099 & 0.500 & 0.446 & 0.053 \\
& BN & 0.014 & 0.142 & 0.520 & 0.776 & 0.076 & 0.028 & 0.164 & 0.531 & 0.728 & 0.095 \\
\cmidrule{2-12}
& Reg+FS & 0.006 & 0.076 & 0.503 & 0.413 & 0.050 & 0.002 & 0.070 & 0.497 & 0.411 & 0.061 \\
& BN+FS & 0.022 & 0.294 & 0.524 & 0.644 & 0.074 & 0.025 & 0.181 & 0.539 & 0.673 & 0.092 \\
\bottomrule[1.5pt]
\end{tabular}}

\label{tab:q1q2_appendix}
\end{table}



\section{Additional Ablation Study}
\paragraph{Can CUBIC’S latent fusion and binary encoding boost model performance in Chinese and HK stock market index prediction?}
As Shown in Table~\ref{tab:q1q2_appendix} Our experiments on Hong Kong and China markets further validate that fusion in latent space and binary encoding classification can independently enhance model performance. For MLP, introducing BN significantly improves model performance: in Hong Kong, IC increases from 0.009 to 0.013, and SR from 0.376 to 0.413; similar improvements are observed in the China market, with IC rising from 0.012 to 0.024 and SR from 0.385 to 0.444. In LSTM models, BN + Single demonstrates clear advantages over Reg + Single when processing constituent stock information: in Hong Kong, IC improves from -0.004 to 0.004; in China, SR substantially increases from -0.346 to 0.596. For Transformer models, the introduction of BN brings notable improvements: in Hong Kong, IC rises from 0.005 to 0.014, and ICLR from 0.046 to 0.142; in China, DA increases from 0.500 to 0.531. When further incorporating feature fusion (FS), model performance gains additional enhancement, with BN+FS achieving an AR of 0.074 and ICLR of 0.294 in the Hong Kong market. These results confirm that the binary classification approach enhances market directional prediction through discrete decision transformation, while feature space fusion improves feature representation learning by aggregating cross-asset features, jointly enhancing the model's overall performance.

\paragraph{How effective is \textsc{Cubic's} digit encoding across market states?} Building upon our previous investigations into \textsc{Cubic}'s effectiveness, we extend our research scope to examine the validity of our weighted cross-entropy loss design. Specifically, we hypothesize that different positions in our binary prediction carry varying levels of importance, which we encode through position-specific weights in the loss function. Additionally, we seek to evaluate how this weighted encoding mechanism performs across different market states, examining its adaptability and effectiveness under varying market conditions. 

As shown in Table~\ref{tab:q3} After adding weights to the base model, the SR in Hong Kong market increased substantially from 0.776 to 0.922, representing an 18.8\% improvement. The US market saw an increase from 0.712 to 0.829, marking a 16.4\% improvement, while the Chinese market rose from 0.728 to 0.882, achieving a 21.2\% improvement. When combined with other components, the effect of weighting becomes even more pronounced - for instance, after adding weights to the BF+Mean model, the SR reached 0.965 in Hong Kong, 1.234 in the US, and 0.961 in China. In the most sophisticated configuration of BF+Mean+DM+Weight, the three markets achieved exceptional performances of 0.977, 1.555, and 1.197 respectively. This consistent pattern of performance enhancement strongly indicates that the position weighting mechanism effectively enhances the model's ability to identify significant price movements. From a theoretical perspective, this improvement stems from the characteristic of binary price encoding where higher bits represent larger price movements. By assigning greater weights to higher positions, the model more accurately captures important market trend changes while effectively suppressing noise effects from lower bits. Cross-market comparison reveals that the weighting strategy performs particularly well in the more volatile US market, reaching a maximum SR of 1.555, further confirming its advantage in capturing significant price movements. Moreover, the fact that significant improvements were achieved across all test markets strongly supports the universality and reliability of this weighting mechanism.

\begin{table*}[ht]
\centering
\caption{The importance of the weight mechanism. The base model is Transformer.}
\setlength{\tabcolsep}{1mm}
\resizebox{0.9\textwidth}{!}{
\begin{tabular}{c||ccccc|ccccc|ccccc}
\toprule[1.5pt]
& \multicolumn{5}{c|}{\textbf{Hong Kong}} &\multicolumn{5}{c|}{\textbf{USA}} & \multicolumn{5}{c}{\textbf{China}}\\
\midrule[1pt]
& IC & ICLR & DA & SR & AR & IC & ICLR & DA & SR & AR & IC & ICLR & DA & SR & AR\\
\midrule[1pt]
BF &  0.014 & 0.142 & 0.520 & 0.776 & 0.076 & 0.025 & 0.450 & 0.531 & 0.712 & 0.085    & 0.028 & 0.164 & 0.531 & 0.728 & 0.095 \\
+Weight & 0.015 & 0.177 & 0.536 & 0.922 & 0.101 & 0.031 & 0.322 & 0.535 & 0.829 & 0.099 & 0.024 & 0.268 & 0.539 & 0.882 & 0.095   \\
\midrule
BF+Mean & 0.021 & 0.220 & 0.523 & 0.834 & 0.106 & 0.031 & 0.373 & 0.559 & 0.785 & 0.102 & 0.034 & 0.317 & 0.559 & 0.809 & 0.083  \\
+Weight & 0.060 & 0.584 & 0.540 & 0.965 & 0.145 & 0.034 & 0.384 & 0.549 & 1.234 & 0.136 & 0.039 & 0.462 & 0.540 & 0.961 & 0.106   \\
\midrule
BF+Trend & 0.037 & 0.242 & 0.528 & 0.919 & 0.101 & 0.036 & 0.311 & 0.552 & 0.783 & 0.110 & 0.038 & 0.257 & 0.560 & 0.964 & 0.104    \\
+Weight & 0.059 & 0.783 & 0.546 & 0.970 & 0.141 & 0.037 & 0.370 & 0.568 & 1.068 & 0.139 & 0.042 & 0.422 & 0.541 & 0.827 & 0.117  \\
\midrule
BF+Mean+DM & 0.036 & 0.258 & 0.531 & 0.860 & 0.135 & 0.038 & 0.401 & 0.558 & 1.350 & 0.149 & 0.053 & 0.584 & 0.565 & 1.329 & 0.121    \\
+Weight & 0.052 & 0.381 & 0.543 & 0.977 & 0.176 & 0.038 & 0.348 & 0.549 & 1.555 & 0.157 & 0.053 & 0.563 & 0.578 & 1.197 & 0.146    \\
\midrule
BF+Trend+DM & 0.036 & 0.371 & 0.531 & 0.828 & 0.158 & 0.040 & 0.471 & 0.566 & 1.210 & 0.143 & 0.052 & 0.552 & 0.568 & 0.923 & 0.127    \\
+Weight &  0.060 & 0.429 & 0.547 & 0.939 & 0.178 & 0.042 & 0.454 & 0.568 & 1.181 & 0.142 & 0.049 & 0.516 & 0.584 & 1.483 & 0.170 \\
\bottomrule[1.5pt]
\end{tabular}}
\label{tab:q3}
\end{table*}

\end{document}